\documentclass[letterpaper,twocolumn,prx,aps,showpacs,superscriptaddress,
floatfix ] {revtex4-1}
\usepackage[latin1]{inputenc}
\usepackage{bm}
\usepackage[usenames]{color}
\usepackage{multirow}
\usepackage{amssymb}
\usepackage{amsbsy}
\usepackage{amsmath}
\usepackage{mathtools}
\usepackage{stmaryrd}
\usepackage{graphicx}
\usepackage{epsfig}
\usepackage{placeins}
\usepackage{ulem}
\usepackage{braket}
\usepackage[colorlinks,linkcolor=blue,citecolor=blue,urlcolor=blue]{hyperref}
\usepackage{soul}

\makeatletter
\begin{document}
\title{Relaxation of dynamically prepared out-of-equilibrium initial states 
within and beyond linear response theory}

\author{Jonas Richter}
\email{jonasrichter@uos.de}
\affiliation{Department of Physics, University of Osnabr\"uck, D-49069 Osnabr\"uck, Germany}

\author{Mats H. Lamann}
\affiliation{Department of Physics, University of Osnabr\"uck, D-49069 Osnabr\"uck, Germany}

\author{Christian Bartsch}
\affiliation{Fakult\"at f\"ur Physik, Universit\"at Bielefeld, D-33615 
Bielefeld, 
Germany}
\affiliation{Department of Physics, University of Osnabr\"uck, D-49069 Osnabr\"uck, Germany}

\author{Robin Steinigeweg}
\email{rsteinig@uos.de}
\affiliation{Department of Physics, University of Osnabr\"uck, D-49069 Osnabr\"uck, Germany}

\author{Jochen Gemmer}
\email{jgemmer@uos.de}
\affiliation{Department of Physics, University of Osnabr\"uck, D-49069 Osnabr\"uck, Germany}

\date{\today}

\begin{abstract}

We consider a realistic nonequilibrium protocol, where a quantum system in 
thermal equilibrium is suddenly subjected to an external force. Due to this 
force, the system is driven out of equilibrium and the expectation values of 
certain observables acquire a dependence on time. Eventually, upon 
switching off the external force, the system unitarily evolves under its own 
Hamiltonian and, as a consequence, the expectation values of observables 
equilibrate towards specific constant long-time values. Summarizing our main 
results, we show that, in systems which violate the eigenstate thermalization 
hypothesis (ETH), this long-time value exhibits an intriguing dependence on the 
strength of the external force. Specifically, for weak external forces, i.e., 
within the linear response regime, we show that expectation values
thermalize to their original equilibrium values, despite the ETH being 
violated. In contrast, for stronger perturbations beyond linear response, 
the quantum system relaxes to some nonthermal value which depends on the 
previous nonequilibrium protocol. While we present theoretical arguments which 
underpin these results, we also numerically demonstrate our 
findings by studying the real-time dynamics of two low-dimensional quantum spin 
models.  

\end{abstract}

\maketitle


\section{Introduction}

Recent years have witnessed an increased interest in the emergence of 
thermodynamic behavior in closed quantum many-body systems 
\cite{polkovnikov2011, gogolin2016, dalessio2016, borgonovi2016}. At the heart 
of this subject 
lies the question if and how an isolated 
system, undergoing solely unitary time evolution, eventually relaxes to some 
long-time steady state which is compliant with the prediction of statistical 
mechanics, i.e., fixed by a few macroscopic parameters only. 

A key approach which has been put forward to answer this question is the 
eigenstate thermalization hypothesis (ETH) \cite{deutsch1991, srednicki1994, 
rigol2005}. The ETH explains 
thermalization on the basis of individual eigenstates and can be formulated as 
an Ansatz about the matrix structure of local observables in the eigenbasis of 
the respective Hamiltonian. Loosely speaking, it states that for 
generic (nonintegrable) quantum systems, the diagonal matrix elements of local 
operators depend smoothly on energy. (When speaking about a 
violation of the ETH in the following, we generally refer to an absence of 
this property.) Given that the ETH is fulfilled, then, 
independent of the specific out-of-equilibrium initial state, the expectation 
values of such operators will always relax to their thermal values prescribed 
by the microcanonical ensemble. 

While it is already hard to proof the validity of the ETH for a given model 
apart from numerical evidence \cite{santos2010, beugeling2014, kim2014, 
mondaini2017, jansen2019,steinigeweg2013}, 
there are also classes of systems which generically violate this Ansatz, with 
integrable and many-body-localized models \cite{basko2006, nandkishore2015} 
being the prime 
examples. On 
the one hand, 
due to the macroscopic number of quasi(local) conservation laws, 
thermalization to standard statistical ensembles is precluded in integrable 
models. Specifically, the long-time steady state in these systems is 
captured in terms of a suitable generalized Gibbs ensemble instead 
\cite{essler2016, vidmar2016}. On the other hand, many-body localization can 
arise in a system with strong disorder. Due to this disorder, transport ceases 
and the system defies thermalization on indefinite time scales 
\cite{abanin2018}.    

Nevertheless, even for systems which violate the ETH 
(i.e.\ eigenstate expectation values of local operators are {\it not smooth}), 
there still exist out-of-equilibrium initial states for which observables 
dynamically equilibrate to their thermal values at long times,
see e.g. \cite{rigol2012, shiraishi2017}. 
In fact, from a 
mathematical point of view, these states even form the majority of all possible 
initial states. Specifically, this statement is also related 
to the notion of {\it typicality} \cite{popescu2006, goldstein2006, 
reimann2007}. 
One central result of {\it typicality} is the fact that the 
overwhelming majority (Haar measure) of quantum states within 
some energy shell yields expectation values of observables very close to the 
full microcanonical ensemble \cite{lloydPhd, gemmer2004}. Or to rephrase, 
quantum states with visible nonequilibrium properties are mathematically rare.
Remarkably, from this set of rare nonequilibrium initial 
states, the majority will, at long times, again exhibit expectation 
values close to the respective equilibrium value, given some mild conditions 
on the dynamics \cite{reimann2008} that are 
entirely unrelated to the ETH \cite{ikeda2011, riera2012, reimann2010, 
reimann2015}.
It is an intriguing and open question whether actual 
(experimentally 
realizable) out-of-equilibrium initial states fall into this larger subset of 
all possible nonequilibrium states, or if the ETH is indeed a physically 
necessary condition for thermalization, see also Refs.\ 
\cite{bartsch2017, depalma2015, reimann2018}. We will explore 
this question by proposing a specific class of initial states which can be 
tuned close to and far away from equilibrium.

We consider a realistic nonequilibrium protocol, where a quantum 
system in thermal equilibrium is suddenly subjected to an external force. Due 
to this force, the system is 
driven out of equilibrium and the expectation values of certain observables
acquire a dependence on time. Eventually, upon switching off the external 
force, 
the system unitarily evolves under its own Hamiltonian and, as a consequence, 
the expectation values of observables equilibrate towards specific 
constant long-time values. Summarizing our main results, we unveil that, in 
systems which violate the ETH, this long-time value exhibits an intriguing 
dependence on the strength of the external force. Specifically, for weak 
external forces, i.e., within the validity regime of linear 
response theory (LRT), we show that expectation values thermalize to their 
original equilibrium values, despite 
the ETH being violated. In contrast, for stronger perturbations beyond linear 
response, the quantum system relaxes to some nonthermal value which depends 
on the previous nonequilibrium protocol. Moreover, for nonintegrable 
systems which obey the ETH, we illustrate that the system 
thermalizes for all initial conditions, both within and beyond the linear 
response regime. Our findings exemplify that (apart from the LRT regime) the 
ETH is indeed a physically necessary condition for initial-state-independent 
relaxation and thermalization in realistic situations.
 
This paper is structured as follows. In Sec.\ \ref{Sec::NES}, 
we introduce and discuss the nonequilibrium protocol which 
is considered in this paper. In Sec.\ \ref{Sec::NumIll}, we then present 
numerical results for two quantum lattice models which corroborate our 
findings. We conclude and summarize in Sec.\ \ref{Sec::Conclu}. 


\section{Nonequilibrium Dynamics}\label{Sec::NES}

\begin{figure}[tb]
 \centering
 \includegraphics[width=0.85\columnwidth]{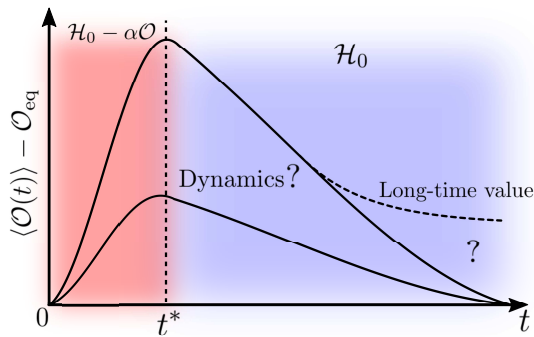}
\caption{(Color online) Sketch of the nonequilibrium setup. For times 
$t < t^\ast$ the system evolves w.r.t.\ a perturbed Hamiltonian ${\cal H}_0 
-\alpha{\cal O}$. Depending on the magnitude of $\alpha$, the expectation value 
$\langle {\cal O}(t)\rangle$ can be driven close to or far away from its 
equilibrium value ${\cal O}_\text{eq}$. For times $t > t^\ast$ the system 
evolves with respect to ${\cal H}_0$. We study how the long-time value $\langle 
{\cal O}(t\to \infty)\rangle$ depends on the perturbation strength $\alpha$.}
 \label{Fig1}
\end{figure}

\subsection{General protocol}

Let us consider a quantum system in thermal equilibrium, i.e., at 
time $t = 0$ it is described by a canonical density matrix, 
\begin{equation}\label{Eq::Rho0}
 \rho(0) = \rho_\text{eq} = \frac{e^{-\beta {\cal H}_0}}{{\cal Z}_0}\ , 
\end{equation}
where ${\cal Z}_0 = \text{Tr}[\exp(-\beta {\cal H}_0)]$ is the partition 
function, $\beta = 1/T$ is the inverse temperature, and ${\cal H}_0$ denotes 
the Hamiltonian of the (unperturbed) quantum system. Next, we consider a 
nonequilibrium protocol where an external static force of strength $\alpha$ is 
suddenly switched on at time $t = 0$, and switched off again at $t = t^\ast$. 
For times $t < t^\ast$, this external force acts on the quantum system, giving 
rise to an additional operator ${\cal O}$ (conjugated to the force) within the 
Hamiltonian. Thus, the full (time-dependent) Hamiltonian ${\cal H}_t$ of the 
nonequilibrium protocol takes on the form
\begin{equation}\label{Eq::Ht}
 {\cal H}_t = \begin{cases} 
                {\cal H}_0 - \alpha {\cal O}\ , & 0 < t < 
t^\ast \\
                {\cal H}_0 \ , & t > t^\ast
               \end{cases}\ .
\end{equation}
The combination of external force and quantum system is considered as being 
isolated from its environment, i.e., the initial equilibrium state $\rho(0)$ 
evolves unitarily in time according to the von-Neumann equation, and $\rho(t)$ 
is given by
\begin{equation}\label{Eq::Rhot}
 \rho(t) = \begin{cases}
             e^{-i({\cal H}_0- \alpha {\cal O}) t} \rho(0) e^{i({\cal H}_0- 
             \alpha {\cal O})t}\ ,& 0 < t < t^\ast \\
             e^{-i {\cal H}_0 (t-t^\ast)} \rho(t^\ast) 
             e^{i{\cal H}_0(t-t^\ast)}\ ,&  t > t^\ast
           \end{cases}\ .
\end{equation}

In the following, we will be interested in the dynamics of the same observable 
${\cal O}$ which is used to perturb the system. While at time $t = 0$, ${\cal 
O}$ takes on its equilibrium value $\langle {\cal O}(0) \rangle = {\cal 
O}_\text{eq}$, the expectation value $\langle {\cal O}(t) \rangle$ acquires a 
time dependence due to the driving by the external force (assuming $[{\cal 
O},{\cal H}_0] \neq 0$). Specifically, for any time $t \geq 0$, $\langle {\cal 
O}(t) \rangle$ reads 
\begin{equation}
 \langle {\cal O}(t) \rangle = \text{Tr}[\rho(t){\cal O}]\ , 
\end{equation}
where $\rho(t)$ is the out-of-equilibrium state in Eq.\ \eqref{Eq::Rhot}.
Clearly, the time dependence of $\langle {\cal O}(t) \rangle$ can be 
manifold. Naively, one might expect a scenario as sketched in Fig.\ \ref{Fig1}. 
For times $t \leq t^\ast$, the expectation value $\langle {\cal O}(t)\rangle$ 
starts increasing with a growth rate depending on the strength of the 
perturbation $\alpha$. The value of $\langle {\cal O}(t)\rangle$ at time $t = 
t^\ast$ is here denoted by 
\begin{equation}
 \langle {\cal O}(t^\ast)\rangle = {\cal O}^\ast\ . 
\end{equation}
Subsequently, for $t > t^\ast$, $\langle {\cal O}(t)\rangle$ evolves w.r.t.\ 
the unperturbed Hamiltonian ${\cal H}_0$ and might relax back to ${\cal 
O}_\text{eq}$, or potentially also to some other long-time value,
\begin{equation}
 \langle {\cal O}(t\to \infty)\rangle = {\cal O}^\infty\ . 
\end{equation}
Note that while the mere process of {\it equilibration} can be 
shown under very little assumptions on ${\cal H}_0, {\cal O}$ 
\cite{reimann2008, short2011}, the present paper is concerned with the  
question of {\it thermalization}. In particular, we address the questions (i) 
how ${\cal O}^\infty$ depends on the strength of the perturbation $\alpha$, and 
(ii) how this dependence is affected by the validity or breakdown of the 
eigenstate thermalization hypothesis.

\subsection{Linear response regime}\label{Sec::LRT_ETH}

Let us now discuss the nonequilibrium protocol outlined above in more detail. 
In fact, in the regime of {\it small} external forces, the 
time dependence of $\langle {\cal O}(t) \rangle$ can be simplified 
considerably. According to linear response theory, the dynamics 
of $\langle {\cal O}(t) \rangle$ in this regime follows as \cite{kubo1991}
\begin{equation}\label{Eq::LRT}
 \langle {\cal O}(t) \rangle - {\cal O}_\text{eq} = \alpha \int_0^{t^\ast} 
\phi(t-t')\ {\text d}t', \
\end{equation}
where we have exploited that the external force $\alpha(t) = \alpha$ 
is constant and only acts for $t\in [0,t^\ast]$.  
Moreover, the function $\phi(t) = -\beta 
\frac{\text{d}}{\text{d}t}(\Delta {\cal O}; \Delta 
{\cal O}(t))$ in Eq.\ \eqref{Eq::LRT} is given 
in terms of a Kubo scalar product \cite{kubo1991},
\begin{equation}
 \phi(t) = -\frac{\text{d}}{\text{d}t}\int_0^\beta \text{Tr}\left[e^{\lambda 
{\cal 
H}_0} \Delta{\cal O}e^{-\lambda {\cal H}_0} \Delta{\cal O}(t) \rho_\text{eq} 
\right] {\text d}\lambda\ \label{Eq::Kubo2},
\end{equation}
where $\Delta {\cal O} =  {\cal O} - {\cal O}_\text{eq}$ and ${\cal O}(t) = 
e^{i\mathcal{H}_0t} {\cal O} e^{-i\mathcal{H}_0t}$. Starting from Eqs.\ 
\eqref{Eq::LRT} and \eqref{Eq::Kubo2}, let us now scrutinize the long-time 
dynamics $\langle {\cal O}(t \to \infty) \rangle$. To this end, we first define
$\chi(t) = \beta (\Delta {\cal O} ; \Delta{\cal O}(t))$, such that $\phi(t) = 
-\text{d} \chi(t)/\text{d}t$ \cite{Note0}. Next, let 
$t_\text{eq}$ be the time for which $\chi(t)$ 
equilibrates, i.e, $\chi(t)$ is essentially time-independent for $t 
> t_\text{eq}$, 
\begin{equation}\label{Eq::ChiConst}
 \chi(t > t_\text{eq}) \approx \overline{\chi} = \text{const}\ .
\end{equation}
Moreover, let $\tau >t_\text{eq}$ be a (long) time 
which is chosen such that $\tau - t_\text{eq}
> t^\ast$. In view of Eq.\ \eqref{Eq::LRT}, the expectation 
value $\langle {\cal O}(\tau)\rangle$ then follows as
\begin{align}
 \langle {\cal O}(\tau)\rangle - {\cal O}_\text{eq} &= \alpha 
\int_0^{t^\ast} 
\phi(\tau - t)\ \text{d}t \label{Eq::Null1}\\ 
 &= \alpha\big[\chi(\tau) - \chi(\tau - t^\ast)\big] \approx 
0\label{Eq::Null2}\ , 
\end{align}
since $\chi(\tau) = \chi(\tau - t^\ast) \approx \overline{\chi}$, 
cf.\ Eq.\ \eqref{Eq::ChiConst}. In particular, for a fixed value of $t^\ast$, 
there 
always exists a time $\tau$ with $\tau - t_\text{eq} > 
t^\ast$ for which Eq.\ \eqref{Eq::Null2} is valid, and 
in the limit $\tau \to \infty$ we can generally write
\begin{equation}\label{Eq::NoStick}
  {\cal O}^\infty - {\cal O}_\text{eq} \approx 0.
\end{equation}
Thus, for small external forces within the validity regime of LRT, the system 
relaxes back to its original equilibrium value. 
Remarkably, this statement only requires equilibration of 
$\chi(t)$, and is therefore expected to hold for a large number of systems [and 
initial states of the form \eqref{Eq::Rho0} and \eqref{Eq::Rhot}], even if the 
ETH is violated.
This is a first important result of the 
present paper.

However, the expression given in Eq.\ \eqref{Eq::LRT} is only valid 
for small external forces, and terms of the order $\alpha^2, \alpha^3, \dots$ 
can become important when $\alpha$ is increased. As a consequence, Eq.\ 
\eqref{Eq::NoStick} can break down, and it is an intriguing question 
how ${\cal O}^\infty$ changes for values of $\alpha$ beyond LRT.  
This transition between small and large values of $\alpha$ will be in the focus 
of our numerical study in the upcoming section. 

\section{Numerical Analysis}\label{Sec::NumIll}

Let us now numerically study the nonequilibrium protocol outlined in Sec.\ 
\ref{Sec::NES}. First, we introduce our models in Sec.\ \ref{Sec::Models}. 
Then, we describe our numerical approach in Sec.\ \ref{Sec::DQT}, before 
presenting our results in Sec.\ \ref{Sec::Results}.    

\subsection{Models}\label{Sec::Models}

\subsubsection{The XXZ chain}\label{Sec::XXZ}

As a first example, we consider the one-dimensional anisotropic Heisenberg 
model (XXZ chain) with periodic boundary conditions. 
The model is described by the Hamiltonian 
\begin{equation}\label{Ham::XXZ}
 {\cal H}_0 = J\sum_{l=1}^L \left(S_l^x S_{l+1}^x + S_l^y S_{l+1}^y + \Delta 
S_l^z S_{l+1}^z \right)\ ,  
\end{equation}
where the $S_l^\mu$, $\mu = x,y,z$ are spin-$1/2$ operators at lattice site 
$l$, $J = 1$ is the antiferromagnetic exchange constant, $L$ is the 
number of lattice sites, and $\Delta \geq 0$ 
denotes the exchange anisotropy in the $z$-direction. 

As an observable for our nonequilibrium protocol, we here choose the spin 
current ${\cal J}$, which can be defined in terms of a lattice continuity 
equation and takes on the well-known form \cite{heidrichmeisner2007},
\begin{equation}\label{Eq::SCur}
 {\cal J} = J\sum_{l=1}^L \left(S_l^x S_{l+1}^y - S_l^y S_{l+1}^x\right)\ . 
\end{equation}
Thus, as outlined in Eqs.\ \eqref{Eq::Ht} and \eqref{Eq::Rhot}, the 
system evolves w.r.t.\ ${\cal H}_0 - \alpha {\cal J}$ for $t < t^\ast$, 
and we study the relaxation of $\langle {\cal J}(t) \rangle$ for long times.
Note that, while a specific force for this particular operator is probably 
difficult to realize in an experiment, this numerical example 
nevertheless nicely illustrates the main results of the present paper. 

The XXZ chain defined in Eq.\ \eqref{Ham::XXZ} is integrable in terms of the 
Bethe Ansatz for all values of $\Delta$ \cite{sutherlandBook}. For the 
particular 
case of $\Delta = 0$, it can be mapped to a model of free spinless fermions 
with ${\cal J}$ being exactly conserved. Moreover, while $[{\cal J}, {\cal 
H}_0]\neq 0$ for all $\Delta \neq 0$, it has been shown that ${\cal 
J}$ is at least partially conserved for anisotropies $\Delta < 1$ 
\cite{zotos1999, prosen2013, urichuk2019}. For 
the purpose of this paper, we therefore choose $\Delta = 0.5$. An explicit 
finite-size scaling, in order to confirm that the ETH is indeed violated for 
this choice of parameters, can be found, e.g., in Ref. \cite{steinigeweg2013}. 

As a comparison, it is furthermore instructive to study the dynamics of ${\cal 
J}$ also in a case where the ETH is valid. To this end, we consider an 
integrability-breaking next-nearest neighbor interaction of strength $\Delta'$, 
i.e., the new Hamiltonian of the system then reads, 
\begin{equation}
 {\cal H}_0' = {\cal H}_0 + J\Delta'\sum_l S_l^z S_{l+2}^z\ . 
\end{equation}
Note that the specific form of 
the spin current \eqref{Eq::SCur} importantly remains unaffected. In 
particular, we here choose $\Delta = \Delta' = 0.5$ for which 
quantum chaos occurs \cite{rigol2010}, and the ETH is 
expected to hold \cite{steinigeweg2013, richter2018_3}.

\subsubsection{The asymmetric spin ladder}\label{Sec::Ladder}

As a second example, we study an asymmetric and anisotropic spin-$1/2$ ladder. 
The Hamiltonian of the spin ladder has a leg part ${\cal H}_\parallel$ 
and a rung part ${\cal H}_\perp$,  
\begin{equation}\label{Eq::Ladder}
 {\cal H}_0 = {\cal H}_\parallel + {\cal H}_\perp\ , 
\end{equation}
where ${\cal H}_\parallel$ essentially consists of two separate XXZ chains, 
cf.\ Eq.\ \eqref{Ham::XXZ}, with different lengths $L_1, L_2$, exchange 
constant $J_\parallel$, and open boundary conditions. Moreover, these two 
chains are then connected according to
\begin{equation}
{\cal H}_\perp = J_\perp \sum_{l=1}^{L_1} S_{l,1}^x S_{l,2}^x +
S_{l,1}^y S_{l,2}^y + \Delta S_{l,1}^z S_{l,2}^z\ , 
\end{equation}
where we have chosen $L_1 < L_2$ without loss of generality. The total 
number of lattice sites is $L = L_1 + L_2$. Based on a finite-size analysis 
of level statistics and of fluctuations of diagonal matrix elements, the 
spin ladder \eqref{Eq::Ladder} has been shown to undergo a transition between a 
chaotic phase and an ETH-violating phase for 
large interchain couplings $J_\perp/J_\parallel \gtrsim 4$ \cite{khodja2015, 
khodja2016}. Since our goal is not to thoroughly 
screen all parameter regimes, but rather to numerically illustrate the 
physical mechanisms discussed in Sec.\ \ref{Sec::NES}, we here choose 
two representative points in parameter space.
Namely, for the chaotic phase we choose $J_\perp = 0.2$, 
whereas for the nonchaotic regime we have $J_\perp = 4.2$. Moreover, we fix 
$J_\parallel = 1$ and $\Delta = 0.1$, cf.\ Refs.\ 
\cite{bartsch2017, khodja2016}.

Furthermore, as an observable, we study the magnetization 
difference between the two legs of the spin ladder, 
\begin{equation}
 {\cal M} = \sum_{l=1}^{L_1} S_{l,1}^z - \sum_{l=1}^{L_2} S_{l,2}^z\ .  
\end{equation}
In particular, this magnetization difference allows for an intuitive 
understanding of our nonequilibrium protocol. Specifically, the external force 
of strength $\alpha$ would correspond to a magnetic field which is directed in 
positive $z$-direction on the first leg, and in negative $z$-direction on the 
second leg. 

\subsection{Dynamical quantum typicality}\label{Sec::DQT}

In order to evaluate time-dependent expectation values 
$\langle {\cal O}(t) \rangle$ for large system sizes (outside the range of 
exact diagonalization), we here rely on an efficient pure-state approach based 
on the concept of dynamical quantum typicality (DQT) \cite{hams2000, 
iitaka2003, sugiura2013, elsayed2013, steinigeweg2014}. Within this concept, a 
{\it single} (randomly drawn) pure quantum state can imitate the properties of 
the full density matrix. Specifically, in order to calculate the expectation 
value $\langle {\cal O}(t) \rangle$, the trace $\text{Tr}[\rho(t) 
{\cal O}]$ is replaced by a simple scalar product,
\begin{equation}\label{Eq::Typical}
 \langle {\cal O}(t) \rangle = \bra{\psi_\beta(t)} {\cal O} \ket{\psi_\beta(t)} 
+ \epsilon\ . 
\end{equation}
Here, $\ket{\psi_\beta(t)}$ denotes the unitarily time-evolved state 
[analogous to Eq.\ \eqref{Eq::Rhot}], and $\ket{\psi_\beta(0)}$ is a
{\it typical} state at inverse temperature $\beta$ \cite{sugiura2013, 
steinigeweg2014}, 
\begin{equation}\label{Eq::Psib}
 \ket{\psi_\beta(0)} = \frac{e^{-\beta {\cal
H}_0/2}\ket{\varphi}}{\sqrt{\bra{\varphi}e^{-\beta {\cal H}_0}\ket{\varphi}}}\ 
,\quad 
 \ket{\varphi} = \sum_{k=1}^d c_k \ket{\varphi_k}\ ,
\end{equation}
where the reference pure state $\ket{\varphi}$ would correspond to
infinite temperature. In particular, the complex coefficients $c_k$ in 
Eq.\ \eqref{Eq::Psib} are randomly drawn from a Gaussian 
distribution with zero mean (Haar measure) \cite{bartsch2009}, and the sum runs 
over the full Hilbert space with dimension $d = 2^L$ and basis 
states $\ket{\varphi_k}$ (in practice, we here choose the Ising basis). Note 
that the statistical error $\epsilon = \epsilon(\ket{\varphi})$ 
in Eq.\ \eqref{Eq::Typical} scales as $\epsilon \propto 1/\sqrt{d_\text{eff}}$, 
where $d_\text{eff} = {\cal Z}_0/e^{-\beta E_0}$ is the effective 
dimension of the Hilbert space, and $E_0$ is the ground-state energy of 
${\cal H}_0$ \cite{hams2000, bartsch2009, elsayed2013, steinigeweg2014, 
steinigeweg2015}. Thus, $\epsilon$ decreases exponentially with system 
size, and the typicality approximation becomes very accurate if $L$ is 
sufficiently large (especially for small values of $\beta$) \cite{Note2}. 
See also Ref.\ \cite{endo2018} for a recent study of linear and 
nonlinear response using typical pure states, as well as Refs.\ 
\cite{bartsch2017, richter2018, richter2018_2, richter2019} for a different but 
related nonequilibrium setup.
\begin{figure}[tb]
 \centering
 \includegraphics[width=0.95\columnwidth]{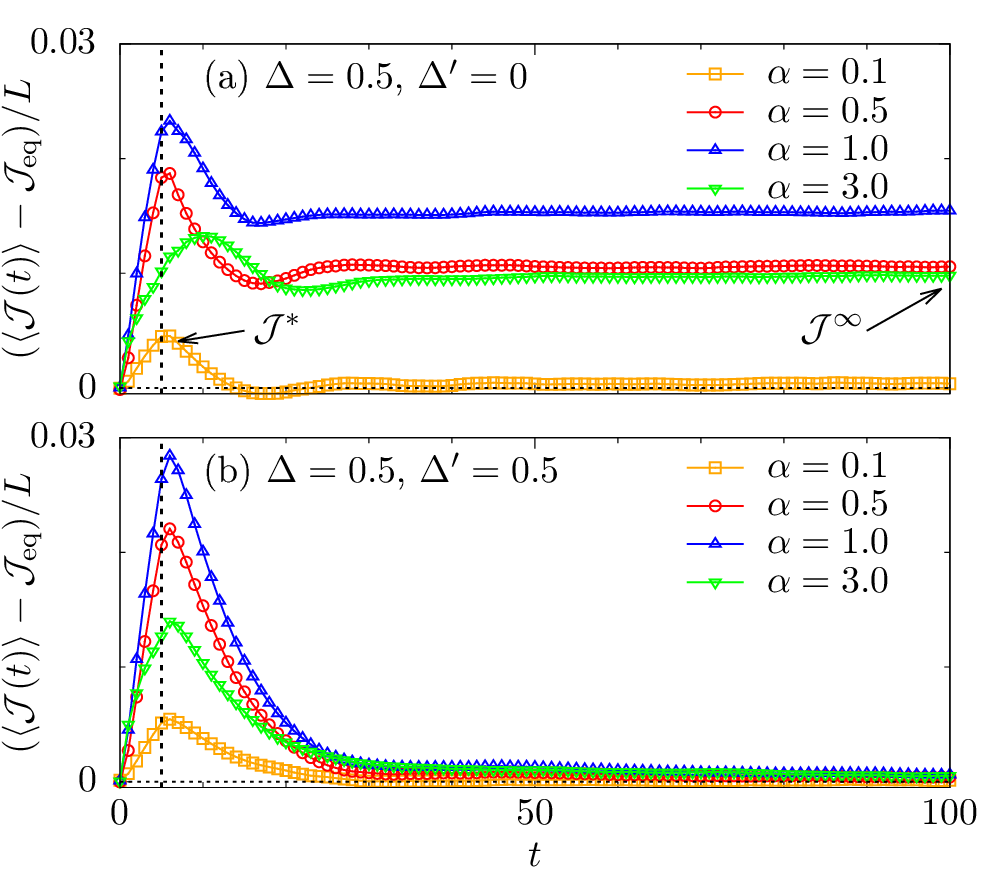}
 \caption{(Color online) (a) Expectation value $(\langle 
{\cal 
J}(t) \rangle -{\cal J}_\text{eq})/L$ for the XXZ chain with $\Delta = 0.5$, 
$\Delta' = 0$. Data is shown for various strengths of the external force 
$\alpha$, which acts up to times $t^\ast = 5$ (as indicated by the dashed 
vertical line). The horizontal dashed line signals zero.
(b) Analogous data as in panel (a), but now for the 
nonintegrable model with $\Delta = \Delta' = 0.5$. The other parameters are $L 
= 24$ and $\beta =1$.}
 \label{Fig2}
\end{figure}

The main numerical advantage of Eq.\ \eqref{Eq::Typical} stems from the fact 
that instead of density matrices, one only has to deal with pure states. 
Particularly, in order to construct the states $\ket{\psi_\beta(0)}$ and 
$\ket{\psi_\beta(t)}$, the exponentials $e^{-\beta \mathcal{H}_0/2}$ or 
$e^{-i\mathcal{H}t}$ can be efficiently evaluated by iteratively solving the 
imaginary- or real-time Schr\"odinger equation, respectively. While various 
sophisticated methods are available for this task, such as Trotter 
decompositions \cite{deReadt2006}, Chebychev polynomials \cite{dobrovitski2003, 
weisse2006}, and Krylov subspace techniques \cite{varma2017}, we here rely on a 
fourth order Runge-Kutta scheme where the discrete time step is chosen short 
enough to guarantee negligible numerical errors \cite{elsayed2013, 
steinigeweg2014}. Such iterator methods, in combination with the sparseness of 
generic few-body operators, enable the treatment of Hilbert-space dimensions 
significantly larger compared to standard exact diagonalization 
\cite{steinigeweg2014, steinigeweg2015, richter2019_1, richter2019_2}. 

\subsection{Results}\label{Sec::Results}

We now present our numerical results. Note that our data are calculated for a  
single temperature $\beta = 1$ only. In particular, we have chosen this 
moderate temperature since it (i) is low enough such that the system can be 
driven out of equilibrium with reasonable effort \cite{Note1}, but (ii) is high 
enough to ensure that finite-size effects and numerical errors are small 
\cite{Note2}. Moreover, while details of the nonequilibrium dynamics can of 
course vary with temperature, the overall picture is expected 
to apply to all values of $\beta$, i.e., there will be a regime of 
sufficiently small $\alpha$ where LRT holds, as well as a 
regime of large $\alpha$ where LRT breaks down. Naturally, the 
notion of {\it small} and {\it large} can depend on the chosen 
temperature $T$. (Note that there can exist cases where LRT breaks down at $T 
= 0$ \cite{gamayun2014})  
\begin{figure}[tb]
 \centering
 \includegraphics[width=0.85\columnwidth]{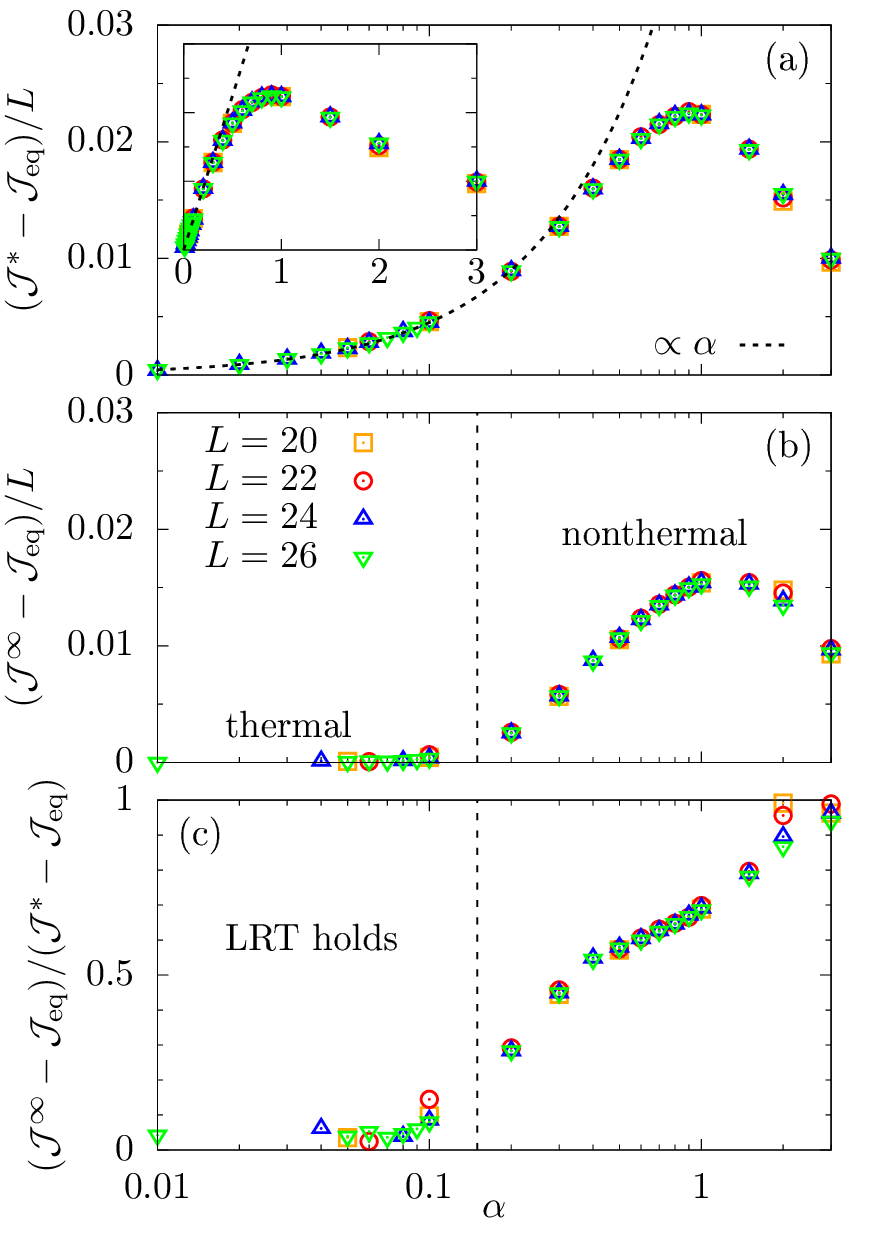}
 \caption{(Color online) (a) $({\cal J}^\ast-{\cal J}_\text{eq})/L$ versus 
perturbation strength $\alpha$. Inset shows same data, but with linear 
$\alpha$-axis. (b) $({\cal J}^\infty-{\cal J}_\text{eq})/L$ versus $\alpha$. 
(c) Ratio $({\cal J}^\infty-{\cal J}_\text{eq})/({\cal 
J}^\ast-{\cal J}_\text{eq})$ versus $\alpha$. Data are shown for different 
system sizes $L = 20,\dots, 26$. Note that ${\cal J}^\infty$ is extracted at 
time $t = 100 \gg t^\ast$. The other parameters are $\Delta = 0.5$, $\Delta' = 
0$, and $\beta = 1$.}
 \label{Fig3}
\end{figure}

\subsubsection{The XXZ chain}

To begin with, we consider the spin current ${\cal J}$ in the XXZ chain as 
introduced in Sec.\ \ref{Sec::XXZ}. In order to get a general impression how 
the dynamics of ${\cal J}$ depends on the strength of the external force, the 
expectation value $\langle {\cal J}(t) \rangle$ is exemplarily shown in 
Figs.\ \ref{Fig2}~(a) and (b) for different values $\alpha = 
0.1, 0.5, 1, 3$, and a single system 
size $L = 24$, both for the integrable as well as the 
nonintegrable model. The nonequilibrium protocol is here designed in such a 
way that 
the external force acts for times $0 < t < 5$, i.e., we have $t^\ast = 5$, as 
indicated by the dashed vertical line.  

First, for short times $t < t^\ast$, we observe in all cases a 
monotonic increase of 
$\langle {\cal J}(t) \rangle$ with time, consistent with the fact that the 
system is driven out of equilibrium. Specifically, comparing data for $\alpha = 
0.1, 0.5$ and $\alpha = 1$, we moreover find that the growth rate of $\langle 
{\cal J}(t) \rangle$ increases with $\alpha$, such that ${\cal J}^\ast = \langle 
{\cal J}(t^\ast)\rangle$ is larger for larger $\alpha$. Quite 
counterintuitively, however, we find that for an even stronger $\alpha = 3$, 
the value of ${\cal J}^\ast$ is actually smaller compared to $\alpha = 0.5, 
1$. In fact, for this large value of $\alpha$, the maximum of 
$\langle {\cal J}(t)\rangle$ in the integrable model [Fig.\ 
\ref{Fig2}~(a)] is shifted to times $t \approx 10$ which is 
considerably beyond $t^\ast$, as if the system does not notice that the external 
force has been already removed. Such a qualitative change in the dynamics 
clearly indicates a transition from linear to nonlinear response when going 
from smaller to larger values of $\alpha$.
\begin{figure}[tb]
 \centering
 \includegraphics[width=0.95\columnwidth]{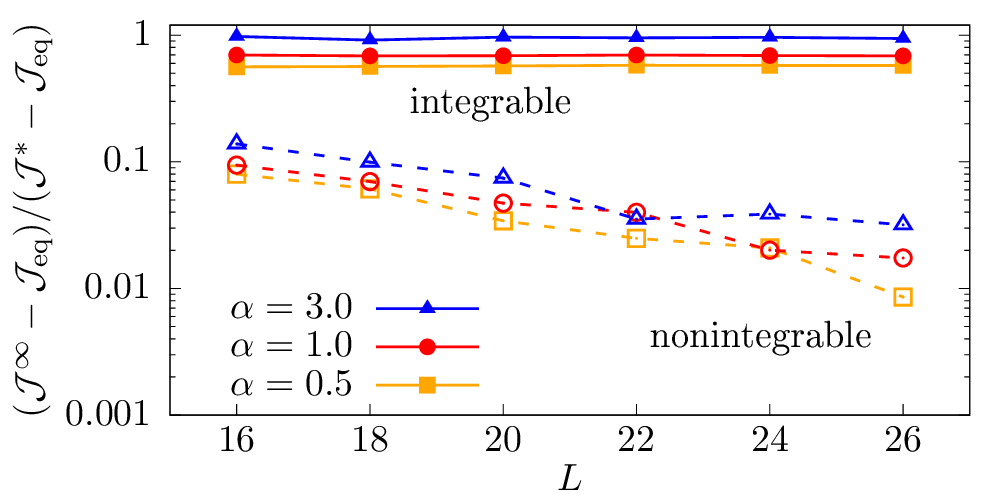}
 \caption{(Color online) Ratio $({\cal J}^\infty-{\cal J}_\text{eq})/({\cal 
J}^\ast-{\cal J}_\text{eq})$ versus system size $L$ for different values of 
$\alpha$ (outside the LRT regime), obtained by diagonalization ($L = 16$) and 
DQT ($L \geq 18$). Data are shown for the integrable model with $\Delta = 
0.5$, $\Delta' = 0$, and the nonintegrable model with $\Delta = \Delta' = 0.5$. 
Since the value of ${\cal J}^\infty$ is extracted at the finite 
time $t = 100$, the data have to be understood as upper bounds to the 
true infinite-time limit (especially for the nonintegrable model)}. We have 
$\beta = 1$ in all cases.
 \label{Fig4}
\end{figure}

Next, concerning the dynamics for $t > t^\ast$ in Fig.\ 
\ref{Fig2}~(a), we observe that after reaching its maximum (approximately at 
$t^\ast$), $\langle 
{\cal J}(t) \rangle$ starts to decrease again, before 
eventually equilibrating to an approximately constant value ${\cal J}^\infty$ at 
long times. While ${\cal J}^\infty-{\cal J}_\text{eq} \approx 0$ in the 
case of $\alpha = 0.1$ (see also further discussion below), we 
find that ${\cal J}^\infty-{\cal J}_\text{eq}$ 
clearly takes on a nonzero value for larger $\alpha$. In the following, we will 
analyze this dependence of ${\cal J}^\infty$ (and in particular the dependence 
of the ratio ${\cal J}^\infty/{\cal J}^\ast$) on the strength of the external 
force $\alpha$ in the integrable model in more detail. 
In contrast, for the nonintegrable model shown in Fig.\ 
\ref{Fig2}~(b), we find that ${\cal J}^\infty-{\cal J}_\text{eq}$ is very small 
and appears to vanish irrespective of the specific value of $\alpha$ (although 
the considered time scales might still be too short).  

In Figs.\ \ref{Fig3}~(a)-(c), the three quantities ${\cal 
J}^\ast-{\cal J}_\text{eq}$, ${\cal 
J}^\infty-{\cal J}_\text{eq}$, and $({\cal J}^\infty-{\cal J}_\text{eq})/({\cal 
J}^\ast-{\cal J}_\text{eq})$ are depicted for integrable 
chains ($\Delta = 0.5$, $\Delta' = 0$) with different system sizes $L \geq 20$ 
and a number of $\alpha$ ranging from  
$\alpha = 0.01$ up to $\alpha = 3$. Note that the $\alpha$-axis has a 
logarithmic scale. First of all, as shown in Fig.\ \ref{Fig3}~(a), we observe a 
linear increase of ${\cal J}^\ast$ for small $\alpha \lesssim 0.2$, 
as expected from linear response theory. (This fact can also be seen in the 
inset of Fig.\ \ref{Fig3}~(a) which has a linear axis.) Moreover, for $\alpha 
\gtrsim 0.2$, deviations from this linear growth become apparent, and for even 
larger $\alpha \gtrsim 1$, one finds that ${\cal J}^\ast$ decreases with 
increasing $\alpha$, consistent with our discussion in the context of Fig.\ 
\ref{Fig2}. As a side remark, while ${\cal J}^\ast$ is not necessarily the 
maximum of $\langle {\cal J}(t)\rangle$, cf.\ Fig.\ 
\ref{Fig2}~(a), the overall findings would be very similar if we plotted this 
maximum instead of ${\cal J}^\ast$. 

Next, Fig.\ \ref{Fig3}~(b) shows the long-time value ${\cal J}^\infty$, which 
is extracted from the real-time dynamics at time $t = 100 \gg t^\ast$. On the 
one hand, for small $\alpha \lesssim 0.2$, we observe that ${\cal 
J}^\infty-{\cal J}_\text{eq} \approx 0$, which is in good agreement with the 
linear regime found in Fig.\ \ref{Fig3}~(a), and consistent with our discussion 
in Sec.\ \ref{Sec::NES}. (The fact that ${\cal 
J}^\infty-{\cal J}_\text{eq}$ is not strictly zero can be explained by (i) the  
finite system size $L$, (ii) the finite time $t = 100$ to extract ${\cal 
J}^\infty$, and (iii) the statistical error $\epsilon$ of the typicality 
approximation.) 
On the other hand, for $\alpha \gtrsim 0.2$, we find 
that ${\cal J}^\infty$ takes on a nonthermal value. Eventually, let us 
emphasize that the data shown in Figs.\ 
\ref{Fig3}~(a) and (b) are normalized to the respective system size $L$, 
resulting in a convincing data 
collapse for all values of $\alpha$ and $L$ shown. This indicates that our 
findings are not just caused by trivial finite-size effects.   

Since both ${\cal J}^\ast-{\cal J}_\text{eq}$ and ${\cal J}^\infty-{\cal 
J}_\text{eq}$ become small for $\alpha \to 0$, it is instructive to study their 
ratio 
\begin{equation}\label{Eq::RR}
{\cal R} = ({\cal 
J}^\infty-{\cal J}_\text{eq})/({\cal J}^\ast-{\cal J}_\text{eq})\ . 
\end{equation}
As can be seen in Fig.\ \ref{Fig3}~(c), this ratio is very small for 
$\alpha \lesssim 0.2$, and drastically changes its behavior for $\alpha 
\gtrsim 0.2$. This clearly confirms our earlier findings from 
Fig.\ \ref{Fig3}~(b). Namely, for small $\alpha$ within the validity regime of 
LRT, ${\cal J}$ relaxes back to its original equilibrium independent of the 
specific out-of-equilibrium 
state. In contrast, for stronger $\alpha$ beyond LRT, ${\cal J}$ 
equilibrates at a nonthermal value ${\cal J}^\infty$, and in particular, this 
${\cal J}^\infty$ clearly depends on the previous nonequilibrium protocol, 
i.e., on the specific value of $\alpha$. This is an important result of the 
present paper. (See also Ref.\ \cite{kennes2017} for similar findings.)

While Fig.\ \ref{Fig3} already shows data for different system sizes $L$, let 
us perform a detailed finite-size scaling for selected values of $\alpha$. In 
this context, it is especially instructive to study how our findings change if 
an integrability-breaking next-nearest neighbor interaction is considered. To 
this end, Fig.\ \ref{Fig4} shows the ratio ${\cal R}$, cf.\ Eq.\ 
\eqref{Eq::RR}, as a function of $L$ for $\alpha = 0.5, 1, 3$ (outside 
the LRT regime). As already discussed above, we find that ${\cal R}$ 
essentially does not exhibit any dependence on system size for the integrable 
(ETH-violating) model. Thus, even in the thermodynamic limit $L \to \infty$, 
the system does not thermalize at long times. In contrast, if we consider the 
nonintegrable model where the ETH holds \cite{steinigeweg2013}, we observe that 
${\cal R}$ clearly decreases with increasing $L$ for all values of $\alpha$ 
shown here, and will likely vanish for $L \to \infty$. This 
exemplifies that, for our realistic nonequilibrium protocol, the ETH is indeed 
a necessary condition for thermalization (at least for $\alpha$ beyond LRT). 
This is another important result. 
\begin{figure}[tb]
 \centering
 \includegraphics[width = 0.95\columnwidth]{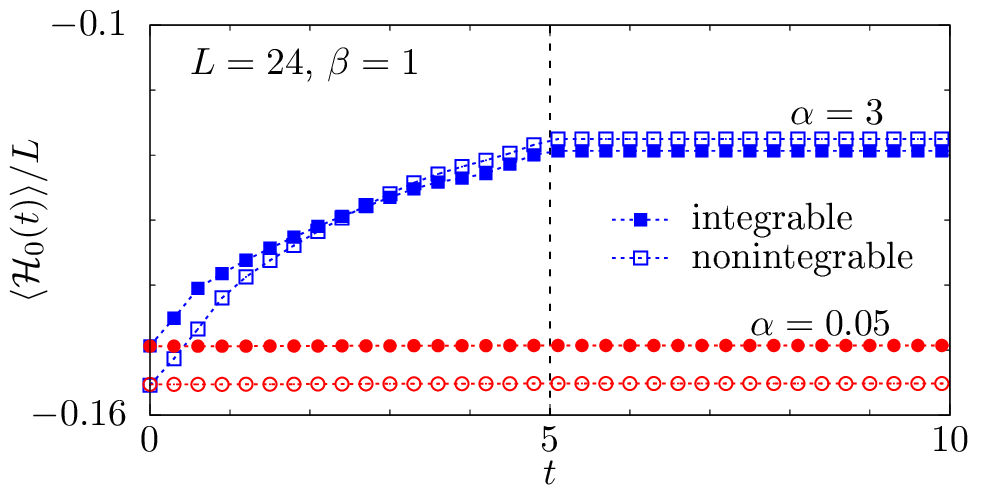}
 \caption{(Color online) Expectation value of the energy density $\langle {\cal 
H}_0(t)\rangle/L$ for the integrable ($\Delta = 0.5$, $\Delta' = 0$, filled 
symbols), and the nonintegrable ($\Delta = \Delta' = 0.5$, open symbols) XXZ 
chain. Data are shown for a small value $\alpha = 0.05$ and 
large value $\alpha = 3$ of the external force. The other parameters 
are $L = 24$ 
and $\beta = 1$.}
 \label{Fig5}
\end{figure}

Eventually, let us note that due to the driving by 
the external force, the system experiences a change of the internal energy 
\cite{mierzejewski2011}. This heating is monitored in Fig.\ \ref{Fig5}, where 
we show $\langle {\cal H}_0(t) \rangle/L$ for the XXZ chain, both for the 
integrable and the nonintegrable model. Moreover, we depict data for 
a small external force 
$\alpha$ within LRT, and a strong external force beyond LRT. On the one hand, 
for a small $\alpha = 0.05$, we observe that $\langle {\cal H}_0(t)\rangle$ 
is essentially constant over the whole time window $t < t^\ast$. (Note that for 
times $t > t^\ast$, $\langle {\cal H}_0(t)\rangle$ is trivially 
time-independent.) On the other hand, for a large $\alpha = 3$, we 
find that $\langle {\cal H}_0(t)\rangle$ monotonically increases, such that 
$\langle {\cal H}_0(t^\ast)\rangle \neq \langle {\cal H}_0(0)\rangle$.
In this context, it is important to stress that the 
{\it nonthermal} long-time value ${\cal J}^\infty$ shown in Fig.\ 
\ref{Fig3}~(b) for $\alpha \gtrsim 0.2$ cannot be explained by this change of 
the internal energy, i.e., it is not just a new thermal value at a different 
effective temperature. In particular, it follows from symmetry considerations 
that the (isolated) XXZ chain cannot carry a nonzero spin current in 
equilibrium, such that ${\cal J}_\text{eq} = 0$ for all energy densities 
\cite{steinigeweg2013}. In fact, the {\it nonthermal} long-time value ${\cal 
J}^\infty$ in Fig.\ \ref{Fig3}~(b) is a direct consequence of the violation of 
the ETH and can be related to the fluctuations of the diagonal matrix elements 
of ${\cal J}$ \cite{steinigeweg2013, richter2018}.
\begin{figure}[tb]
 \centering
 \includegraphics[width=0.95\columnwidth]{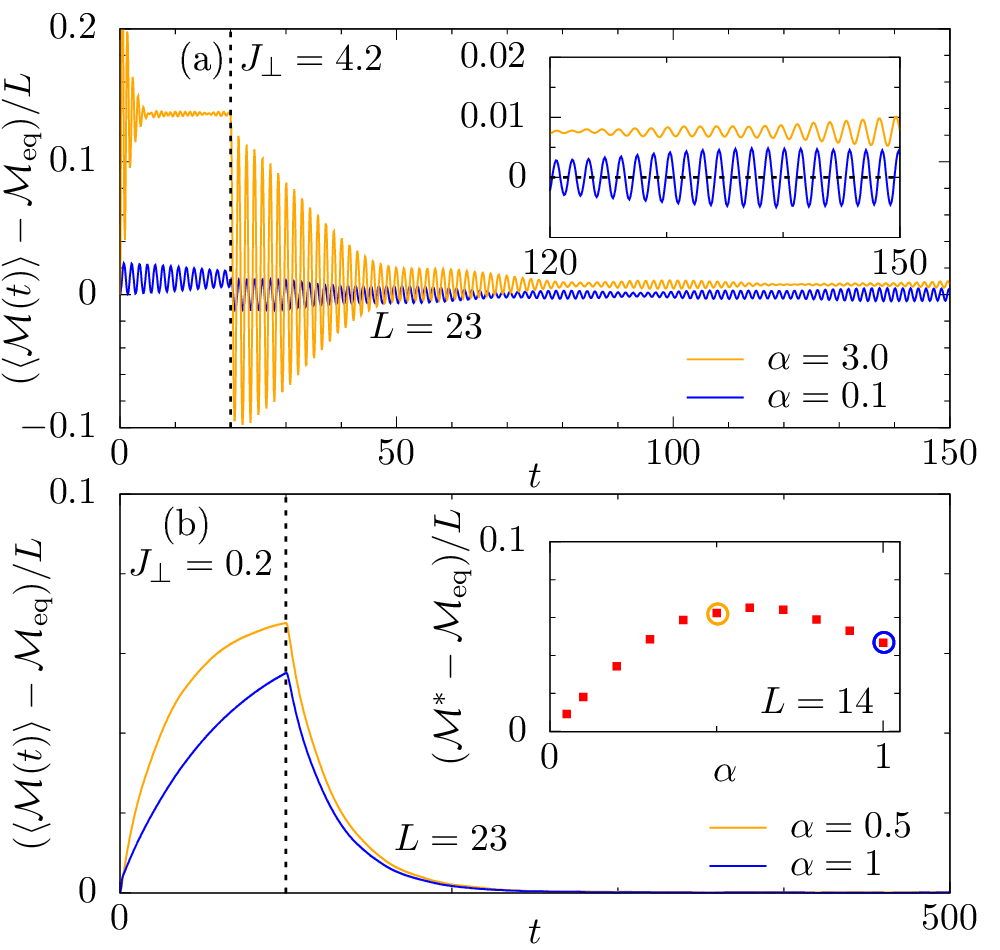}
 \caption{(Color online) (a) Expectation value of the 
magnetization difference $(\langle {\cal M}(t)\rangle-{\cal M}_\text{eq})/L$ 
in the strong-coupling regime with $J_\perp = 4.2$, 
exemplarily shown for $\alpha = 0.1, 3$ and system size $L_1 = 
8$, $L_2 = 15$. Inset shows same data, but only for the time window $120 < 
t < 150$. (b) Analogous data as in panel (a), but now for the 
chaotic regime with $J_\perp = 0.2$ and $\alpha = 0.5, 1$. The inset shows the 
dependence of ${\cal M}^\ast$ on $\alpha$, obtained by diagonalization for $L_1 
= 5$, $L_2 = 9$ [see Fig.\ \ref{Fig7}~(a) for data at $J_\perp = 4.2$]. The 
small circles indicate the values of ${\cal M}^\ast$ and $\alpha$ which 
correspond to the two curves shown in the main panel. The vertical dashed lines 
in (a) and (b) indicate the duration of the external perturbation $t^\ast = 20, 
100$. The other parameters are, $J_\parallel = 1$, $\Delta = 0.1$, and 
$\beta =1$.}
 \label{Fig6}
\end{figure}

\subsubsection{The asymmetric spin ladder}
\begin{figure}[tb]
 \centering
 \includegraphics[width = 0.9\columnwidth]{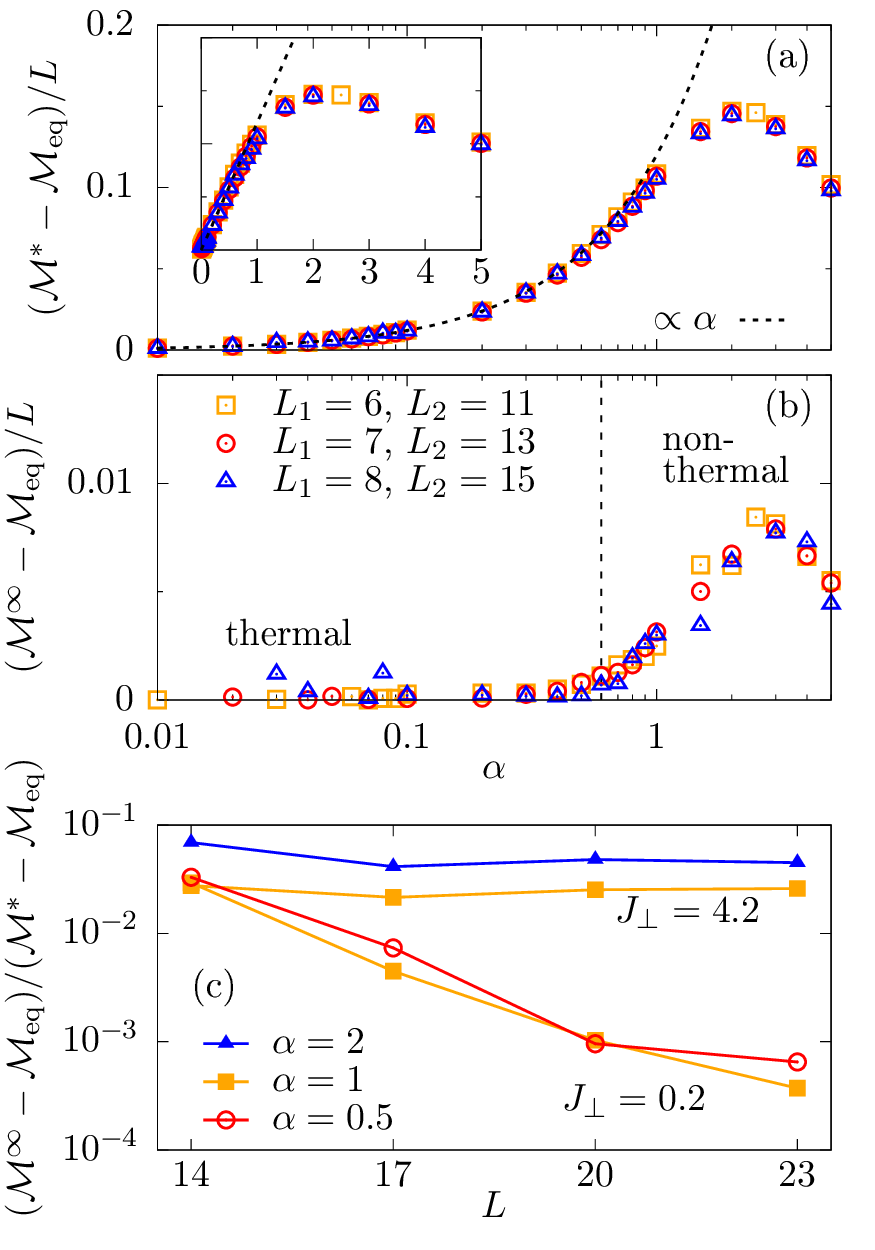}
 \caption{(Color online) (a) $({\cal M}^\ast-{\cal M}_\text{eq})/L$ versus 
$\alpha$ for $L = 17, 20, 23$ [see key in (b) for details] and 
$J_\perp = 4.2$. Inset shows same data, but with linear $\alpha$-axis. (b) 
$({\cal M}^\infty-{\cal M}_\text{eq})/L$ versus $\alpha$ for $J_\perp = 4.2$. 
(c) Ratio $({\cal M}^\infty-{\cal M}_\text{eq})/({\cal 
M}^\ast-{\cal M}_\text{eq})$ versus system size $L$ for $\alpha = 0.5,1,2$. 
Data is shown for both, strong interchain coupling $J_\perp = 4.2$ 
where the ETH is violated, and weak $J_\perp = 0.2$ where the system is 
chaotic. The other parameters are 
$J_\parallel = 1$, $\Delta = 0.1$, and $\beta = 1$.}
 \label{Fig7}
\end{figure}

To corroborate our findings further, let us now also study the asymmetric spin 
ladder introduced in Sec.\ \ref{Sec::Ladder}. In Figs.\ 
\ref{Fig6}~(a) and (b), we again exemplarily depict the nonequilibrium 
dynamics of the magnetization 
difference $\langle {\cal M}(t)\rangle$ for different perturbation strengths 
$\alpha$, both for the ETH-violating regime ($J_\perp = 4.2$) 
and the chaotic regime ($J_\perp = 0.2$).
In case of the former [Fig.\ \ref{Fig6}~(a)], we find that 
the magnetization difference 
exhibits a sudden drop at $t = t^\ast$. Moreover, due to the rather strong rung 
coupling, $\langle {\cal M}(t) \rangle$ shows pronounced oscillations which 
also persist up to the longest time $t = 150$ considered. Due to this 
oscillatory behavior, we extract both ${\cal M}^\ast$ and ${\cal M}^\infty$ as 
an average over a suitably chosen time window. In contrast, for 
the chaotic regime in Fig.\ \ref{Fig6}~(b), the dynamics is considerably slower 
such that we choose a larger $t^\ast = 100$ and extract ${\cal M}^\infty$ 
around $t \approx 500$. 

Next, Figs.\ \ref{Fig7}~(a) and \ref{Fig7}~(b) show $({\cal M}^\ast-{\cal 
M}_\text{eq})/L$ and $({\cal M}^\infty-{\cal M}_\text{eq})/L$ 
versus $\alpha$ for different system sizes $L$ and $J_\perp = 
4.2$. Analogous to our discussion of the spin current in Fig.\ \ref{Fig3}~(a), 
we again find a regime of small $\alpha \lesssim 0.6$ where ${\cal M}^\ast$ 
grows linearly with $\alpha$. 
Furthermore, for $\alpha \gtrsim 2$ we also observe the counterintuitive 
phenomenon that ${\cal M}^\ast$ decreases although the external force becomes 
stronger.

Concerning the long-time value shown in Fig.\ \ref{Fig7}~(b), we find 
${\cal M}^\infty-{\cal M}_\text{eq} \approx 0$ for $\alpha 
\lesssim 0.6$, as well as a monotonic growth of ${\cal M}^\infty$ for $\alpha 
\gtrsim 0.6$. Thus, although the overall effect is considerably weaker in the 
case of the spin ladder (see also \cite{Note3}), Fig.\ \ref{Fig7}~(b) confirms 
our previous findings from Figs.\ \ref{Fig2} to \ref{Fig4}. In particular, we 
again can clearly identify two separate regimes, i.e., a first regime for weak 
$\alpha$ where $\langle {\cal M}(t) \rangle$ takes on its thermal value at long 
times, and a second regime for larger $\alpha$ where LRT breaks down and ${\cal 
M}^\infty$ is nonthermal. 

Since qualitatively similar, we have omitted in Fig.\ \ref{Fig7} the analogous 
panel (c) compared to Fig.\ \ref{Fig4}. Instead, Fig.\ 
\ref{Fig7}~(c) presents a finite-size scaling of ${\cal R}_{\cal M} = ({\cal 
M}^\infty-{\cal M}_\text{eq})/({\cal M}^\ast-{\cal M}_\text{eq})$ for 
different values of the external force $\alpha$. In particular, we also 
show data for the chaotic region of the parameter space with smaller rung 
couping $J_\perp = 0.2$, where the ETH is expected to apply \cite{bartsch2017, 
khodja2015, khodja2016}. [Note however, that the LRT regime is found to be  
smaller in the chaotic regime, see inset of Fig. \ref{Fig6}~(b)]. On the one 
hand, for strong ${\cal J}_\perp = 4.2$, we find that ${\cal R}_{\cal M}$ is 
practically independent of the system size. On the other hand, for ${\cal 
J}_\perp = 0.2$, we find that ${\cal R}_{\cal M}$ decreases approximately 
exponentially for increasing $L$. Thus, the validity of the ETH in the chaotic 
regime ensures thermalization at long times, independent of the specific 
initial state.

\section{Conclusion}\label{Sec::Conclu}

To summarize, we have studied a particular type of nonequilibrium protocol 
where a quantum system in thermal equilibrium is suddenly subjected to an 
external force which drives the system out of equilibrium. Eventually, 
this external force is switched off again, and the system evolves under its own 
(unperturbed) Hamiltonian.

As main results, we have shown that, in 
systems which violate the ETH, the long-time value of observables exhibits an 
intriguing dependence on the strength of the external force. Specifically, for 
weak external forces, i.e., within the linear response regime, we unveiled 
that expectation values thermalize to their original equilibrium values, 
despite the ETH being violated. In contrast, for stronger perturbations beyond 
linear response, the quantum system relaxes to some nonthermal value which 
depends on the previous nonequilibrium protocol. 

We have substantiated our results by numerically studying the real-time 
dynamics of observables in two low-dimensional quantum lattice models: (i) the 
spin current in the one-dimensional XXZ model, and (ii) the magnetization 
difference between the two legs in an asymmetric spin ladder. In particular, we 
have employed an efficient pure-state approach in order to study large systems, 
and to demonstrate that our findings do not depend on system size.  
In this context, we have also demonstrated that in the case of a nonintegrable 
model, the system relaxes back to thermal equilibrium (also for 
far-from-equilibrium initial states).
\begin{figure}[tb]
 \centering
 \includegraphics[width = 0.9\columnwidth]{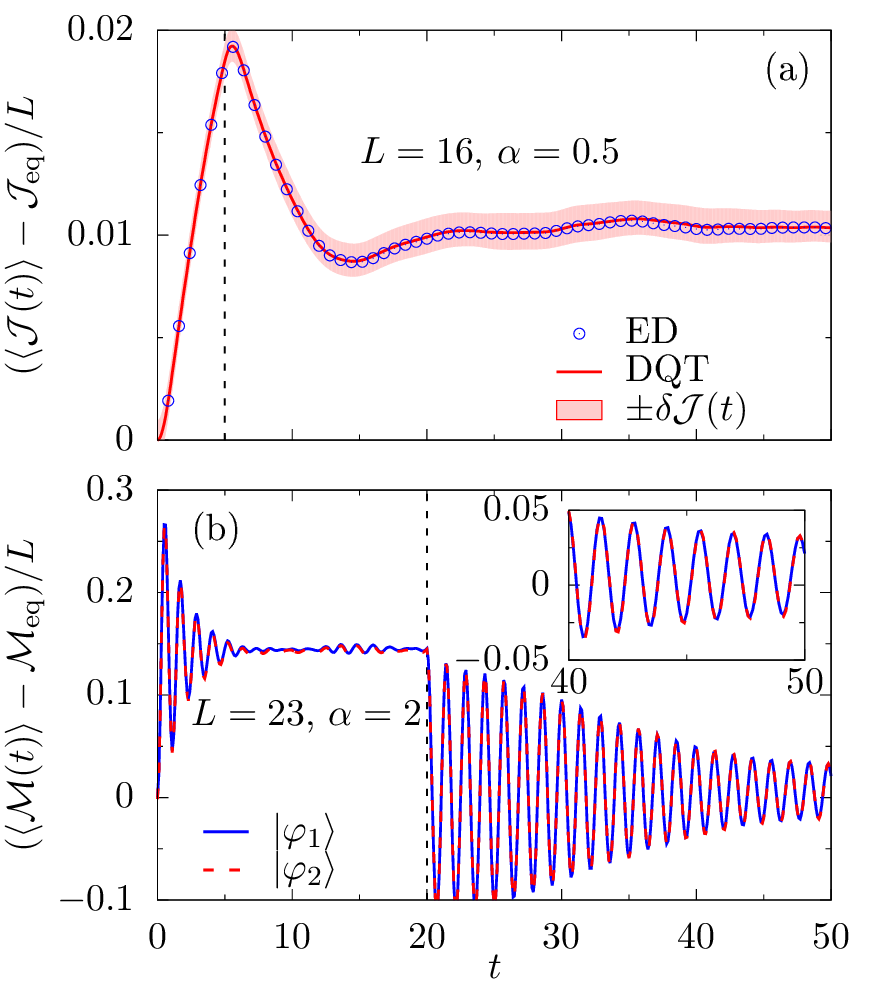}
 \caption{(Color online) (a) Comparison of $\langle {\cal J}(t)\rangle$ 
obtained by the typicality approach, cf.\ Eq.\ \eqref{Eq::Typical}, and by 
exact diagonalization for $L = 16$. DQT data are averaged over $N = 100$ 
samples, and the shaded area indicates sample-to-sample fluctuations, cf.\ 
Eq.\ \eqref{Eq::StS}. Note that the error of the mean scales as $\delta 
\overline{{\cal J}(t)} \propto \delta {\cal J}(t)/\sqrt{N}$. (b) $\langle {\cal 
M}(t)\rangle$ obtained from two different realizations of the pure state 
$\ket{\varphi}$, cf.\ Eq.\ \eqref{Eq::Psib}, for a ladder with 
$L_1 = 8$, $L_2 = 15$. We have 
$\beta = 1$ in all cases.}
 \label{Fig8}
\end{figure}

On the one hand, our findings exemplify that the ETH is indeed a 
physically necessary condition for initial-state-independent 
relaxation and thermalization in realistic situations. On the other hand, and 
almost paradoxically, our nonequilibrium protocol at the same time exhibits the 
intriguing property that systems can thermalize for initial states within the 
LRT regime, despite the ETH being violated. 

Promising directions of research include, e.g., the consideration of other 
time-dependent external perturbations, a more thorough investigation of the 
dependence on temperature, as well as the study of many-body localized systems 
within this nonequilibrium protocol.  

\subsection*{Acknowledgements}

This work has been funded by the Deutsche Forschungsgemeinschaft (DFG) - 
Grants No.\ 397107022 (GE 1657/3-1), No.\ 397067869 (STE 
2243/3-1), No.\ 355031190 - within the DFG Research Unit FOR 2692. 


\appendix

\section{Accuracy of the pure-state approach}\label{Sec::Accu}

In order to demonstrate that dynamical quantum typicality [Eq.\ 
\eqref{Eq::Typical}] indeed provides an accurate numerical approach to study 
nonequilibrium dynamics, Fig.\ \ref{Fig8}~(a) shows a comparison of $\langle 
{\cal J}(t) \rangle$ with exact diagonalization data for a small system of size 
$L = 16$. One clearly observes that both methods agree convincingly with each 
other for all times shown here. In particular, the DQT data are averaged 
over $N$ different random realizations of the pure state $\ket{\varphi}$, cf.\ 
Eq.\ \eqref{Eq::Psib}, and the shaded area indicates the standard deviation of 
sample-to-sample fluctuations \cite{richter2018},
\begin{equation}\label{Eq::StS}
 \delta {\cal J}(t) = \left[ \sum_{n=1}^N \frac{\langle {\cal J}(t) 
\rangle^2_{(n)}}{N} - \left( \sum_{n=1}^N \frac{\langle {\cal J}(t) 
\rangle_{(n)}}{N} \right)^2 \right]^{\tfrac{1}{2}} \ .
\end{equation}  
The error of the mean scales as $\delta \overline{{\cal J}(t)} \propto \delta 
{\cal J}(t)/\sqrt{N}$, and is well-controlled for the choice of $N = 100$ used 
here. As outlined below Eq.\ \eqref{Eq::Psib}, the accuracy of the pure-state 
approach is expected to improve even further for increasing Hilbert-space 
dimension. Therefore, averaging becomes less important for increasing $L$, and 
the data for $L \geq 20$ shown in Figs.\ \ref{Fig2} to \ref{Fig4} essentially 
represents the exact dynamics. Note that data for $L \geq 22$ in Fig.\ 
\ref{Fig4} is calculated from a single pure state $N = 1$ only. Note further 
that the data in Fig.\ \ref{Fig7} has been obtained by averaging over $N = 300$ 
($L = 17$), $N = 100$ ($L = 20$), and $N = 1$ ($L = 23$) states. We expect the 
small fluctuations in Fig.\ \ref{Fig7} to vanish if $N$ is increased further. 

Another convenient means to demonstrate the smallness of the statistical error 
$\epsilon$ is the direct comparison of data resulting from two different 
instances of the typical state $\ket{\varphi}$. Such a comparison is shown in 
Fig.\ \ref{Fig8}~(b) for the magnetization difference $\langle {\cal M}(t) 
\rangle$ in a spin ladder with $L = 23$ sites. One observes that the data 
resulting from $\ket{\varphi_1}$ and $\ket{\varphi_2}$ coincide very well with 
each other, illustrating that Eq.\ \eqref{Eq::Typical} indeed provides a 
reliable tool to obtain quantum many-body dynamics for large system sizes.  



\begin{thebibliography}{199}

\bibitem{polkovnikov2011}
A. Polkovnikov, K. Sengupta, A. Silva, and M. Vengalattore, 
Rev. Mod. Phys. {\bf 83}, 863 (2011).

\bibitem{gogolin2016}
C. Gogolin and J. Eisert,
Rep. Prog. Phys. {\bf 79}, 056001 (2016).

\bibitem{dalessio2016}
L. D'Alessio, Y. Kafri, A. Polkovnikov, and M. Rigol,
Adv. Phys. {\bf 65}, 239 (2016).

\bibitem{borgonovi2016}
F. Borgonovi, F. M. Izrailev, L. F. Santos, and V. G. 
Zelevinsky, Phys. Rep. {\bf 626}, 1 (2016).

\bibitem{deutsch1991}
J. M. Deutsch, 
Phys. Rev. A {\bf 43}, 2046 (1991). 

\bibitem{srednicki1994}
M. Srednicki, 
Phys. Rev. E {\bf 50}, 888 (1994).

\bibitem{rigol2005}
M. Rigol, V. Dunjko, and M. Olshanii, 
Nature {\bf 452}, 854 (2008). 

\bibitem{santos2010}
L. F. Santos and M. Rigol, 
Phys. Rev. E {\bf 82}, 031130 (2010). 

\bibitem{beugeling2014}
W. Beugeling, R. Moessner, and M. Haque, 
Phys. Rev. E {\bf 89}, 042112 (2014). 

\bibitem{kim2014}
H. Kim, T. N. Ikeda, D. A. Huse, 
Phys. Rev. E {\bf 90}, 052105 (2014). 

\bibitem{mondaini2017}
R. Mondaini and M. Rigol, 
Phys. Rev. E {\bf 96}, 012157 (2017).

\bibitem{jansen2019}
D. Jansen, J. Stolpp, L. Vidmar, and F. Heidrich-Meisner, 
Phys. Rev. B {\bf 99}, 155130 (2019). 

\bibitem{steinigeweg2013}
R. Steinigeweg, J. Herbrych, and P. Prelov\v{s}ek, 
Phys. Rev. E {\bf 87}, 012118 (2013).

\bibitem{basko2006}
D. M. Basko, I. L. Aleiner, and B. L. Altshuler, 
Ann. Phys. {\bf 321}, 1126 (2006).

\bibitem{nandkishore2015}
R. Nandkishore and D. A. Huse,
Annu. Rev. Condens. Matter Phys. {\bf 6}, 15 (2015).

\bibitem{essler2016}
F. H. L. Essler and M. Fagotti,
J. Stat. Mech. {\bf 2016}, 064002 (2016).

\bibitem{vidmar2016}
L. Vidmar and M. Rigol,
J. Stat. Mech. {\bf 2016}, 064007 (2016).

\bibitem{abanin2018}
D. A. Abanin, E. Altman, I. Bloch, and M. Serbyn, 
Rev. Mod. Phys. {\bf 91}, 021001 (2019).

\bibitem{rigol2012}
M. Rigol and M. Srednicki, 
Phys. Rev. Lett. {\bf 108}, 110601 (2012). 

\bibitem{shiraishi2017}
N. Shiraishi and T. Mori, 
Phys. Rev. Lett. {\bf 119}, 030601 (2017).

\bibitem{popescu2006}
S. Popescu, A. J. Short, and A. Winter, 
Nat. Phys. {\bf 2}, 754 (2006). 

\bibitem{goldstein2006}
S. Goldstein, J. L. Lebowitz, R. Tumulka, and N. Zangh\`i, 
Phys. Rev. Lett. {\bf 96}, 050403 (2006). 

\bibitem{reimann2007}
P. Reimann,  
Phys. Rev. Lett. {\bf 99}, 160404 (2007).

\bibitem{lloydPhd}
S. Lloyd, 
Ph.D. Thesis, The Rockefeller University (1988), Chapter 3, 
arXiv:1307.0378.

\bibitem{gemmer2004}
J. Gemmer, M. Michel, and G. Mahler,
{\it Quantum Thermodynamics} (Springer, Berlin, 2004). 

\bibitem{reimann2008}
P. Reimann, 
Phys. Rev. Lett. {\bf 101}, 190403 (2008).

\bibitem{ikeda2011}
T. N. Ikeda, Y. Watanabe, and M. Ueda, 
Phys. Rev. E {\bf 84}, 021130 (2011). 

\bibitem{riera2012}
A. Riera, C. Gogolin, and J. Eisert, 
Phys. Rev. Lett. {\bf 108}, 080402 (2012). 

\bibitem{reimann2010}
P. Reimann, 
New. J. Phys. {\bf 12}, 055027 (2010). 

\bibitem{reimann2015}
P. Reimann, 
Phys. Rev. Lett. {\bf 115}, 010403 (2015).

\bibitem{bartsch2017}
C. Bartsch and J. Gemmer,
EPL (Europhys. Lett.) \textbf{118}, 10006 (2017).

\bibitem{depalma2015}
G. De Palma, A. Serafini, V. Giovannetti, and M. Cramer,
Phys. Rev. Lett. {\bf 115}, 220401 (2015). 

\bibitem{reimann2018}
P. Reimann, 
Phys. Rev. Lett. {\bf 120}, 230601 (2018). 

\bibitem{short2011}
A. J. Short, 
New J. Phys. {\bf 13}, 053009 (2011). 

\bibitem{kubo1991}
R. Kubo, M. Toda, and N. Hashitsume,
{\it Statistical Physics II: Nonequilibrium Statistical Mechanics}, 
Solid-State Sciences {\bf 31} (Springer, Berlin, 1991).

\bibitem{Note0}
The definition of $\chi(t)$ is only fixed up to a constant offset, which is 
however irrelevant for our considerations.

\bibitem{heidrichmeisner2007}
F. Heidrich-Meisner, A. Honecker, and W. Brenig, 
Eur. Phys. J. Spec. Top. \textbf{151}, 135 (2007).

\bibitem{sutherlandBook}
B. Sutherland, 
{\it Beautiful Models: 70 Years of Exactly Solved Quantum Many-body Problems} 
(World Scientific Publishing, Singapore, 2004). 

\bibitem{zotos1999}
X. Zotos, 
Phys. Rev. Lett. {\bf 82}, 1764 (1999). 

\bibitem{prosen2013}
T. Prosen and E. Ilievski, 
Phys. Rev. Lett. {\bf 111}, 057203 (2013).

\bibitem{urichuk2019}
A. Urichuk, Y. Oez, A. Kl\"umper, and J. Sirker, 
SciPost Phys. {\bf 6}, 005 (2019). 

\bibitem{rigol2010}
M. Rigol and L. F. Santos, 
Phys. Rev. A {\bf 82}, 011604(R) (2010). 

\bibitem{richter2018_3}
J. Richter, F. Jin, H. De Raedt, K. Michielsen, J. Gemmer, and R. Steinigeweg, 
Phys. Rev. B {\bf 97}, 174430 (2018). 

\bibitem{khodja2015}
A. Khodja, R. Steinigeweg, and J. Gemmer, 
Phys. Rev. E {\bf 91}, 012120 (2015). 

\bibitem{khodja2016}
A. Khodja, D. Schmidtke, and J. Gemmer, 
Phys. Rev. E {\bf 93}, 042101 (2016). 

\bibitem{hams2000}
A. Hams and H. De Raedt, 
Phys. Rev. E {\bf 62}, 4365 (2000). 

\bibitem{iitaka2003} 
T. Iitaka and T. Ebisuzaki,
Phys. Rev. Lett. {\bf 90}, 047203 (2003).

\bibitem{sugiura2013}
S. Sugiura and A. Shimizu,
Phys. Rev. Lett. {\bf 111}, 010401 (2013).

\bibitem{elsayed2013}
T. A. Elsayed and B. V. Fine,
Phys. Rev. Lett. {\bf 110}, 070404 (2013).

\bibitem{steinigeweg2014}
R. Steinigeweg, J. Gemmer, and W. Brenig,
Phys. Rev. Lett. \textbf{112}, 120601 (2014).

\bibitem{bartsch2009}
C. Bartsch and J. Gemmer, 
Phys. Rev. Lett. {\bf 102}, 110403 (2009). 

\bibitem{steinigeweg2015}
R. Steinigeweg, J. Gemmer, and W. Brenig,
Phys. Rev. B {\bf 91}, 104404 (2015).

\bibitem{Note2}
For more details on the accuracy of our pure-state approach, see 
Appendix \ref{Sec::Accu}. 

\bibitem{endo2018}
H. Endo, C. Hotta, and A. Shimizu, 
Phys. Rev. Lett. {\bf 121}, 220601 (2018). 

\bibitem{richter2018}
J. Richter, J. Herbrych, and R. Steinigeweg, 
Phys. Rev. B {\bf 98}, 134302 (2018). 

\bibitem{richter2018_2}
J. Richter, J. Gemmer, and R. Steinigeweg, 
Phys. Rev. E {\bf 99}, 050104(R) (2019).

\bibitem{richter2019}
J. Richter and R. Steinigeweg, 
Phys. Rev. E {\bf 99}, 012114 (2019).

\bibitem{deReadt2006}
H. De Raedt and K. Michielsen, in 
\textit{Handbook of  Theoretical and Computational Nanotechnology}
(American Scientific Publishers, Los Angeles, 2006).

\bibitem{dobrovitski2003}
V. V. Dobrovitski and H. De Raedt, 
Phys. Rev. E {\bf 67}, 056702 (2003).

\bibitem{weisse2006}
A. Wei\ss e, G. Wellein, A. Alvermann, and H. Fehske, 
Rev. Mod. Phys. {\bf 78}, 275 (2006). 

\bibitem{varma2017}
V. K. Varma, A. Lerose, F. Pietracaprina, J. Goold, and A. Scardicchio, 
J. Stat. Mech. {\bf 2017} 053101 (2017). 

\bibitem{richter2019_1}
J. Richter and R. Steinigeweg, 
Phys. Rev. B {\bf 99}, 094419 (2019). 

\bibitem{richter2019_2}
J. Richter, F. Jin, L. Knipschild, J. Herbrych, H. De Raedt, K. Michielsen, J. 
Gemmer, and R. Steinigeweg, 
Phys. Rev. B {\bf 99}, 144422 (2019). 

\bibitem{Note1}
As becomes evident from Eqs.\ \eqref{Eq::LRT} and \eqref{Eq::Kubo2}, a smaller 
value of $\beta$ leads to a smaller response of the system for fixed value of 
$\alpha$.

\bibitem{gamayun2014}
O. Gamayun, O. Lychkovskiy, and V. Cheianov, 
Phys. Rev. E {\bf 90}, 032132 (2014). 

\bibitem{kennes2017}
D. M. Kennes, J. C. Pommerening, J. Diekmann, C. Karrasch, and V. Meden, 
Phys. Rev. B {\bf 95}, 035147 (2017). 

\bibitem{mierzejewski2011}
M. Mierzejewski, J. Bon\v{c}a, and P. Prelov\v{s}ek,
Phys. Rev. Lett. {\bf 107}, 126601 (2011).

\bibitem{Note3}
In fact, the relaxation towards a nonthermal value in the spin 
ladder is considerably more pronounced in a different nonequilibrium setup, see 
Ref.\ \cite{bartsch2017}.



\end{thebibliography}
\end{document}